\documentclass[twocolumn,astrosymb]{aastex631}

\usepackage{amsmath}
\shorttitle{LAEs at z$\sim$7 from LAGER}
\shortauthors{Harish et al.}

\begin{document}

\title{New spectroscopic confirmations of Lyman-$\alpha$ emitters at z $\sim$ 7 from the LAGER survey}

\author{Santosh Harish}
\affil{School of Earth and Space Exploration, Arizona State University, Tempe, AZ 85287, USA}
\email{santosh.harish@asu.edu}

\author{Isak G. B. Wold}
\affil{Astrophysics Science Division, NASA Goddard Space Flight Center, 8800 Greenbelt Road, Greenbelt, Maryland, 20771, USA}

\author{Sangeeta Malhotra}
\affil{Astrophysics Science Division, NASA Goddard Space Flight Center, 8800 Greenbelt Road, Greenbelt, Maryland, 20771, USA}

\author{James Rhoads}
\affil{Astrophysics Science Division, NASA Goddard Space Flight Center, 8800 Greenbelt Road, Greenbelt, Maryland, 20771, USA}

\author{Weida Hu}
\affil{CAS Key Laboratory for Research in Galaxies and Cosmology, Department of Astronomy, University of Science and Technology of China, Hefei, Anhui 230026, People's Republic of China}
\affil{School of Astronomy and Space Science, University of Science and Technology of China, Hefei 230026, People's Republic of China}

\author{Junxian Wang}
\affil{CAS Key Laboratory for Research in Galaxies and Cosmology, Department of Astronomy, University of Science and Technology of China, Hefei, Anhui 230026, People's Republic of China}
\affil{School of Astronomy and Space Science, University of Science and Technology of China, Hefei 230026, People's Republic of China}

\author{Zhen-ya Zheng}
\affil{CAS Key Laboratory for Research in Galaxies and Cosmology, Shanghai Astronomical Observatory, Shanghai 200030, People's Republic of China}

\author{L. Felipe Barrientos}
\affil{Instituto de Astrof\'{i}sica, Facultad de F\'{i}sica, Pontificia Universidad Cat\'{o}lica de Chile, Santiago, Chile}

\author{Jorge Gonz\'{a}lez-L\'{o}pez}
\affil{N\'{u}cleo de Astronomía de la Facultad de Ingenier\'{i}a y Ciencias, Universidad Diego Portales, Av. Ej\'{e}rcito Libertador 441, Santiago, Chile}
\affil{Instituto de Astrof\'{i}sica, Facultad de F\'{i}sica, Pontificia Universidad Cat\'{o}lica de Chile, Santiago, Chile}

\author{Lucia A. Perez}
\affil{School of Earth and Space Exploration, Arizona State University, Tempe, AZ 85287, USA}

\author{Ali Ahmad Khostovan}
\affil{Astrophysics Science Division, NASA Goddard Space Flight Center, 8800 Greenbelt Road, Greenbelt, Maryland, 20771, USA}

\author{Leopoldo Infante}
\affil{Las Campanas Observatory, Carnegie Institution of Washington, Casilla 601, La Serena, Chile}

\author{Chunyan Jiang}
\affil{CAS Key Laboratory for Research in Galaxies and Cosmology, Shanghai Astronomical Observatory, Shanghai 200030, People's Republic of China}

\author{Crist\'{o}bal Moya-Sierralta}
\affil{Instituto de Astrof\'{i}sica, Facultad de F\'{i}sica, Pontificia Universidad Cat\'{o}lica de Chile, Santiago, Chile}

\author{John Pharo}
\affil{School of Earth and Space Exploration, Arizona State University, Tempe, AZ 85287, USA}

\author{Francisco Valdes}
\affil{National Optical Astronomy Observatory, 950 N. Cherry Avenue, Tucson, AZ 85719, USA}

\author{Huan Yang}
\affil{Las Campanas Observatory, Carnegie Institution of Washington, Casilla 601, La Serena, Chile}



\begin{abstract}
We report spectroscopic confirmations of 15 Lyman-alpha galaxies at $z\sim7$, implying a spectroscopic confirmation rate of $\sim$80\% on candidates selected from LAGER (Lyman-Alpha Galaxies in the Epoch of Reionization), which is the largest (24 deg$^2$) survey aimed at finding Lyman-alpha emitters (LAEs) at $z\sim7$ using deep narrow-band imaging from DECam at CTIO.  LAEs at high-redshifts are sensitive probes of cosmic reionization and narrow-band imaging is a robust and effective method for selecting a large number of LAEs. In this work, we present results from the spectroscopic follow-up of LAE candidates in two LAGER fields, COSMOS and WIDE-12, using observations from Keck/LRIS.  We report the successful detection of Ly$\alpha$ emission in 15 candidates (11 in COSMOS and 4 in WIDE-12 fields). Three of these in COSMOS have matching confirmations from a previous LAGER spectroscopic follow-up and are part of the overdense region, LAGER-$z7$OD1. Additionally, two candidates that were not detected in the LRIS observations have prior spectroscopic confirmations from Magellan. Including these, we obtain a spectroscopic confirmation success rate of $\sim$$80$\% for LAGER LAE candidates. Apart from Ly$\alpha$, we do not detect any other UV nebular lines in our LRIS spectra; however, we estimate a 2$\sigma$ upper limit for the ratio of N\textsc{v}/Ly$\alpha$, $f_{NV}/f_{Ly\alpha} \lesssim 0.27$, which implies that ionizing emission from these sources is mostly dominated by star formation. Including confirmations from this work, a total of 33 LAE sources from LAGER are now spectroscopically confirmed. LAGER has more than doubled the sample of spectroscopically confirmed LAE sources at $z\sim7$.

\end{abstract}

\keywords{}


\section{Introduction} \label{sec:intro}
Epoch of Reioniization (EoR) marks a critical phase in the cosmic history when neutral hydrogen (HI) in the intergalactic medium (IGM) was ionized by intense UV radiation from sources such as star-forming galaxies and active galactic nuclei (AGN), soon after the dark ages. Recent efforts including Cosmic Microwave Background (CMB) observations by \emph{Planck} \citep{Planck2016} and high-redshift quasar observations \cite[e.g.][]{Fan2006} suggest that the universe was highly ionized by $z\sim6$; however, a detailed picture of the reionization process is still lacking, given significant challenges in observing the earliest sources in the universe. Among several observational probes that exist for studying reionization, Ly$\alpha$ emission from star-forming galaxies is one of the most powerful and sensitive tracer of HI in the IGM \cite[e.g.][]{Malhotra2004,Malhotra2006,Stark2011,Pentericci2011,Dijkstra2014,Zheng2017,Konno2018}. Ly$\alpha$ photons undergo resonant scattering in the presence of HI and therefore provide a reliable estimate of the HI fraction as the universe ionized. 

Over the last two decades, a number of narrow-band (NB) imaging surveys have successfully selected large samples of Ly$\alpha$ emitters (LAEs) up to $z\sim6$ \cite[e.g.][]{Rhoads2003,Rhoads2004,Malhotra2004,Ouchi2005,Hu2010,Zheng2016}. Some recent studies have found overdensities of Ly$\alpha$ \cite[e.g.]{Zheng2017,Castellano2018,Hu2019} and Lyman-break galaxies at $z\sim7$ \cite[e.g.][]{Trenti2012,Castellano2016} which might indicate presence of large ionized regions. However, at $z\gtrsim7$, attenuation of Ly$\alpha$ photons by the increasingly neutral IGM has resulted in fewer LAE candidates \cite[e.g.][]{Fontana2010,Treu2013,Pentericci2014,Mason2019} and shown a signifcant drop in terms of the Ly$\alpha$ luminosity function \cite[e.g.][]{Ota2010,Shibuya2012,Konno2014}, compared to lower redshifts.

Lyman-Alpha Galaxies in the Epoch of Reionization (LAGER) is an ongoing, large NB survey searching for LAEs at $z\sim7$, using the Dark Energy Camera (DECam) on the 4m Blanco telescope at Cerro Tololo Inter-American Observatory (CTIO), Chile \citep{Zheng2017}. LAGER employs a custom designed NB filter, NB964, with a central wavelength of 9642 \AA.

In this paper, we present follow-up spectroscopic observations of several LAE candidates at $z \sim 7$ in COSMOS and WIDE-12 fields from the LAGER survey. Details of LRIS observations and the data reduction process is presented in Section \ref{sec:obs}. The resultant spectra and LAE identifications are shown in Section \ref{sec:results}, followed by a summary in Section \ref{sec:summary}.

\section{Observations and Data Reduction} \label{sec:obs}

\subsection{Candidate LAE selection}
Using the narrow-band filter NB964 \citep{Zheng2019} -- with a central wavelength of 9642\AA\ and FWHM $\sim$ 90\AA~ -- on the Dark Energy Camera (DECam) mounted on the Blanco 4m telescope at CTIO, we have obtained 47.25 hours narrow-band exposure in the COSMOS field, reaching a 5$\sigma$ detection limit of 25.2 mag (2\arcsec aperture), and in WIDE-12 field, we have obtained 27.9 hours of narrow-band exposure reaching a 5$\sigma$ detection limit of 24.7 mag (1.8\arcsec aperture), as part of the LAGER survey. In COSMOS, LAE candidates at $z \sim 7$ were selected based on narrow-band flux excess ($y$ - NB964 $>$ 0.8) and non-detections in the bluer broad-band imaging (HSC $g, r, i, z$). The final sample contains 49 LAE candidates (for more details, see \citealt{Hu2019}). Similar approach was followed for the WIDE-12 region resulting in a sample of 50 LAE candidates \citep{Wold2021}.

\begin{deluxetable}{ccccc}
	\tablenum{1}
	\tablecaption{Keck/LRIS observation summary \label{tab:lris_obs}}
	\tablewidth{0pt}
	\tablehead{
		\colhead{Mask ID} & \colhead{Observation} & \colhead{No.\ of exposures} &  \colhead{$t_{exp}^{total}$} & \colhead{Seeing}  \\
		\colhead{} & \colhead{Date} & \colhead{$\times$ $t_{exp}^{single}$ (s)}  & \colhead{(h)} & \colhead{}
	}
	\startdata
	WIDE12-19 & Mar 5, 2019 & 10$\times$900 + 1$\times$720 & 2.7 & $0.7\arcsec$\\
	COSMOS-A-19 & Mar 5, 2019 & 10$\times$900 + 5$\times$600 & 3.3 & $0.7\arcsec$\\
	COSMOS-B-19 & Mar 5, 2019 & 7$\times$720 & 1.4 & $0.7\arcsec$\\
	WIDE12-20 & Jan 30, 2020 & 16$\times$720 + 1$\times$600 & 3.36 & $0.6\arcsec$-$0.7\arcsec$\\
	COSMOS-20 & Jan 31, 2020 & 15$\times$720 & 3 & $0.5\arcsec$-$0.7\arcsec$\\
	\enddata
\end{deluxetable}

\subsection{LRIS spectroscopy} \label{sec:lris}
We observed 21 LAE candidates using the Low Resolution Imaging Spectrometer (LRIS) on the Keck I telescope \citep{Oke1995,Rockosi2010} over one night in 2019 (March 5) and two half-nights in 2020 (January 30-31). With a beamsplitter in place, the red and blue cameras of LRIS can be operated simultaneously, providing a combined spectral coverage from 3200\AA\ to 1 $\mu$m with a field-of-view of 6\arcmin$\times$7.8\arcmin. Given the expected redshifts of our candidate LAEs, we focus here on observations obtained by the red camera, which uses a mosaic of two LBNL 2k $\times$ 4k CCD detectors, with a spatial resolution of 0.135\arcsec/pixel. Our setup included the 400/8500 grating and a multi-slit mask. With our typical 1\arcsec slit width, this setup yields a spectral resolution of $\sim$7\AA. Each of our slitmasks in COSMOS and WIDE-12 fields included the $z \sim 7$ LAE candidates, foreground emission-line candidates, and alignment stars.

The 2019 observing run included three masks (2 in COSMOS and 1 in WIDE-12), covering a total of 15 LAE candidates, and the 2020 run included two masks (one each in COSMOS and WIDE-12), covering a total of 6 LAE candidates. Weather conditions were excellent in both runs with a seeing of $0.7\arcsec$ on March 5, 2019, and $0.5\arcsec-0.7\arcsec$ on January $30-31$, 2020. The observation summary including number of exposures and total exposure time per mask is provided in Table \ref{tab:lris_obs}.

\begin{deluxetable*}{ccccccc}
	\tablenum{2}
	\tablecaption{Properties of LAE candidates observed using Keck/LRIS \label{tab:lager_spec}}
	\tablewidth{0pt}
	\tablehead{
		\colhead{ID} & \colhead{Mask} & \colhead{R.A.} & \colhead{Dec.} & \colhead{z$_{spec}$} &  \colhead{NB964} & \colhead{log L$_{Ly\alpha}$$^a$}  \\
		\colhead{} & & \colhead{(J2000)} & \colhead{(J2000)} & \colhead{}  & \colhead{(mag)} & \colhead{(erg s$^{-1}$)}
	}
	\startdata
	& & Confirmations& & &\\
	\hline
	J095933+014320 & COSMOS-20 & 09:59:33.26 & +01:43:20.42 & 6.923 & 24.82 $\pm$ 0.21 & 42.84 $^{+0.08} _{-0.12}$\\
	J095934+014406 & COSMOS-20 & 09:59:34.80 & +01:44:06.96 & 6.937 & 25.18 $\pm$ 0.23 & 42.57 $^{+0.12} _{-0.19}$\\
	J095954+013938 & COSMOS-20 & 09:59:54.37 & +01:39:38.33 & 6.930 & 25.0 $\pm$ 0.2 & 42.73 $^{+0.09} _{-0.12}$\\
	J095956+023259 & COSMOS-A-19 & 09:59:56.04 & +02:32:59.99 & 6.924 & 25.02 $\pm$ 0.2 & 42.77 $^{+0.08} _{-0.12}$\\
	J100000+023219 & COSMOS-A-19 & 10:00:00.43 & +02:32:19.27 & 6.926 & 24.43 $\pm$ 0.14 & 42.99 $^{+0.06} _{-0.08}$\\
	J100001+023402 & COSMOS-A-19 & 10:00:01.84 & +02:34:02.03 & 6.920 & 24.57 $\pm$ 0.21 & 42.8 $^{+0.14} _{-0.19}$\\
	J100012+023047 & COSMOS-A-19 & 10:00:12.95 & +02:30:47.26 & 6.929 & 23.61 $\pm$ 0.15 & 43.31 $^{+0.07}_{-0.08}$\\
	J100327+020851$^b$ & COSMOS-B-19 &  10:03:27.99 & +02:08:51.28 & 6.915 & 24.73 $\pm$ 0.23 & 43.05 $^{+0.09}_{-0.12}$\\
	J100332+020925$^b$ & COSMOS-B-19 & 10:03:32.71 & +02:09:25.11 & 6.898 & 24.33 $\pm$ 0.16 & 43.04 $^{+0.06} _{-0.1}$\\
	J100339+020747$^b$ & COSMOS-B-19 & 10:03:39.32 & +02:07:47.22 & 6.936 & 24.91 $\pm$ 0.22 & 42.69 $^{+0.14} _{-0.2}$\\
	J100339+020954 &  COSMOS-B-19 & 10:03:39.12 & +02:09:54.42 & 6.920 & 25.49 $\pm$ 0.25 & 42.55 $^{+0.11} _{-0.17}$\\	
	J120412-003157 & WIDE12-20 & 12:04:12.68 & -00:31:57.94 & 6.935 & 24.12 $\pm$ 0.07 & 43.02 $^{+0.13}  _{-0.18}$\\
	J120406-002921 & WIDE12-20 & 12:04:06.51 & -00:29:21.40 & 6.907 & 24.44 $\pm$ 0.11 & 42.93 $^{+0.18}  _{-0.32}$\\
	J120356-002648 & WIDE12-20 & 12:03:56.77 & -00:26:48.24 & 6.896 & 23.95 $\pm$ 0.06 & 43.17 $^{+0.11}  _{-0.14}$\\
	J120637-001438 & WIDE12-19 & 12:06:37.34 & -00:14:38.95 & 6.929 & 24.25 $\pm$ 0.08 & 43.02 $^{+0.14}  _{-0.21}$\\
	\hline
	&&Interlopers&&&\\
	\hline
	J120636+000959 & WIDE12-19 & 12:06:36.53 & +00:09:59.07 & 0.925 & 24.21 $\pm$ 0.07 & - \\ 
	J120641-001047 &  WIDE12-19 & 12:06:41.19 & -00:10:47.30 & 0.921 & 23.99 $\pm$ 0.06 & - \\
	\hline
	&&Non-confirmations&&&\\
	\hline
	J100016+022847 & COSMOS-A-19 & 10:00:16.94 & +02:28:47.36 & - & 23.66 $\pm$ 0.13 & 43.25 $^{+0.07}_{-0.09}$\\
	J100333+020719$^b$ & COSMOS-B-19 & 10:03:33.51 & +02:07:19.80 & - & 24.61 $\pm$ 0.2 & 42.94 $^{+0.07}_{-0.09}$\\
	J100334+020909 & COSMOS-B-19 & 10:03:34.63 & +02:09:09.58 & - & 24.87 $\pm$ 0.21 & 42.82 $^{+0.08} _{-0.12}$\\
	J100337+020736$^b$ & COSMOS-B-19 & 10:03:37.34 & +02:07:36.65 & - & 24.81 $\pm$ 0.18 & 42.86 $^{+0.06} _{-0.1}$\\	
	\enddata
	\tablecomments{$^a$Photometrically measured Ly$\alpha$ luminosities using NB964 and HSC $y$-band photometry where Ly$\alpha$ luminosities were derived assuming the NB964 filter's central wavelength as the line wavelength.\\
	$^b$ In \citealt{Hu2021}, J100332+020925, J100327+020851, J100339+020747, J100333+020719 and J100337+020736 are known, respectively, by the following IDs: LAE-4, LAE-6, LAE-19, LAE-17 and LAE-18.}
\end{deluxetable*}

\subsection{Data Reduction} \label{sec:redux}
We reduced our LRIS observations using an open-source, Python-based data reduction pipeline called \texttt{PypeIt} \citep{Prochaska2020:joss,Prochaska2020:zen}. The reduction pipeline consists of a semi-automated script that applies a list of algorithms to each of the raw exposures. All exposures of the same type (such as arc, tilt) are combined together to construct master calibration frames. For each science exposure, PypeIt applies the standard reduction techniques including slit edge tracing, wavelength calibration based on arc exposures, flat-field correction and sky subtraction. Typically, PypeIt performs object extraction and b-spline sky subtraction, jointly. In the end, the pipeline performs both boxcar and optimal extraction to generate the 1D and 2D science spectra. 

In order to achieve a higher signal-to-noise, we combined the extracted 1D and 2D spectra for each object in each mask using the coadd script, \emph{pypeit\_coadd\_2dspec}. Since our targets are expected to be faint sources with little to no continuum, we chose to perform manual extraction of our sources by providing the spatial-spectral pixel pair for each object in the PypeIt reduction file.

\section{Results} \label{sec:results}
\begin{figure*}
	\begin{tabular}{lll}
		\vspace{-0.5em} \hspace{-3.75em}\includegraphics[width=2.4in]{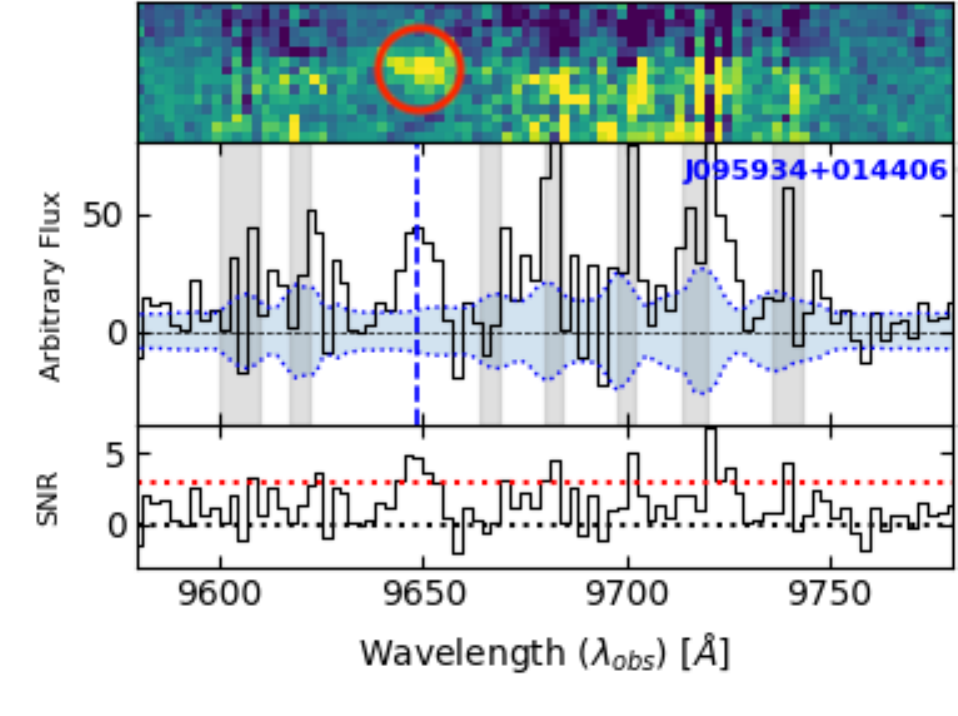} & \hspace{-1.em}\includegraphics[width=2.4in]{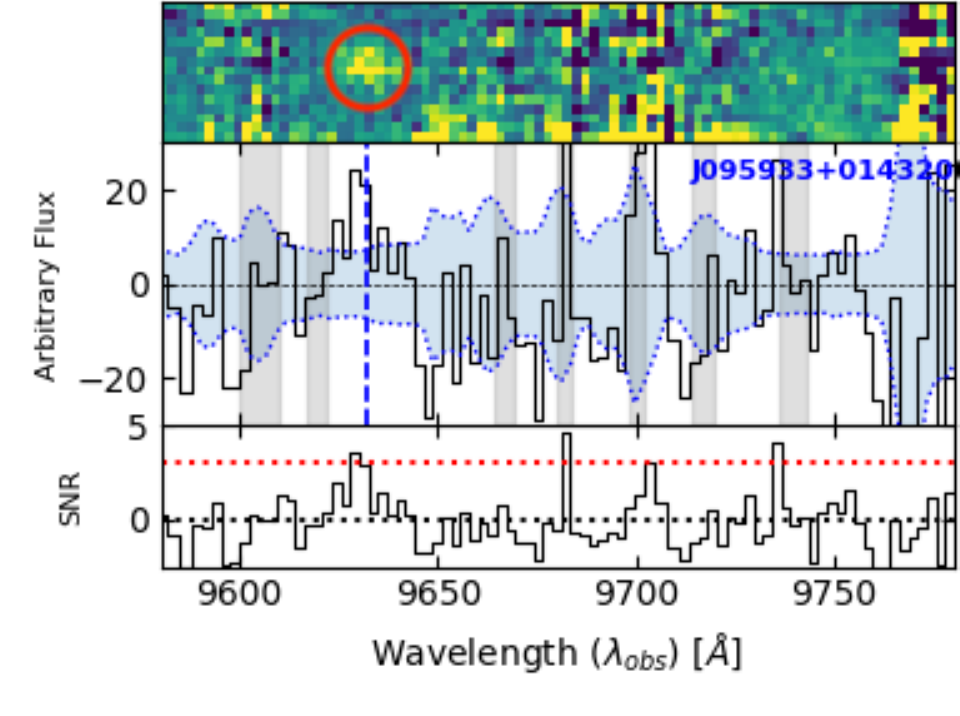} & \hspace{-1.em}\includegraphics[width=2.4in]{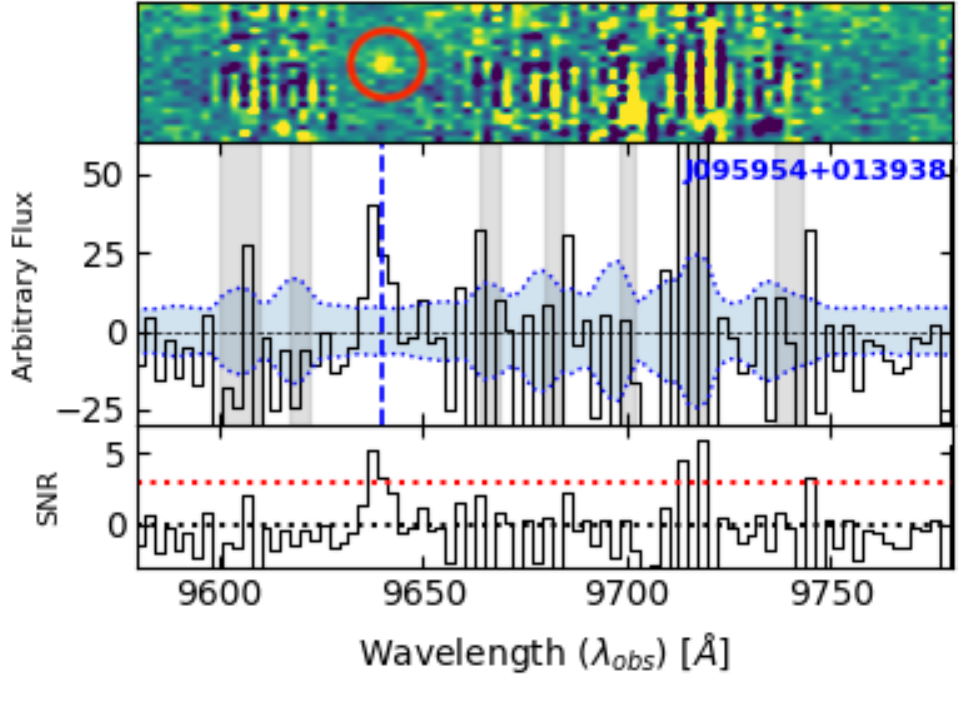}\\ \vspace{-0.5em}
		\hspace{-3.75em}\includegraphics[width=2.4in]{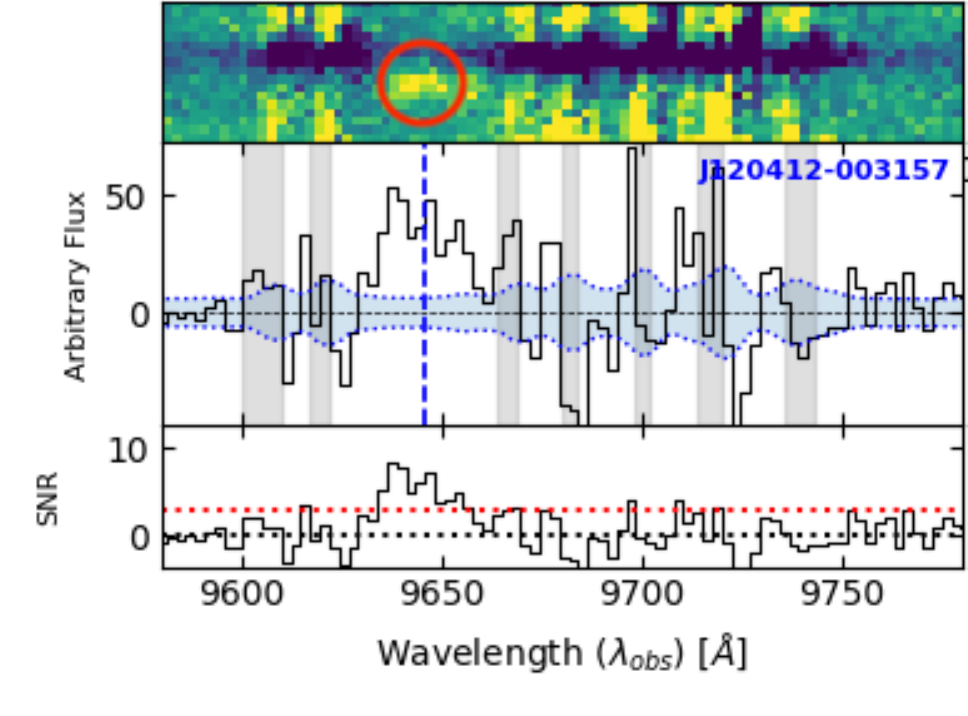} &  \hspace{-1.em}\includegraphics[width=2.4in]{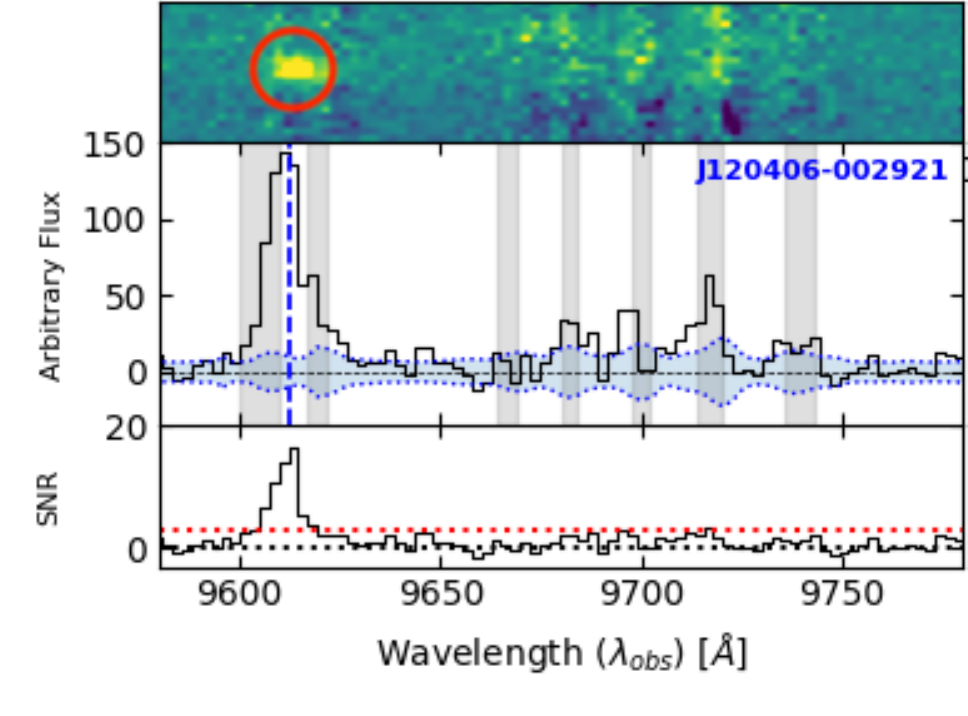} & \hspace{-1.em}\includegraphics[width=2.4in]{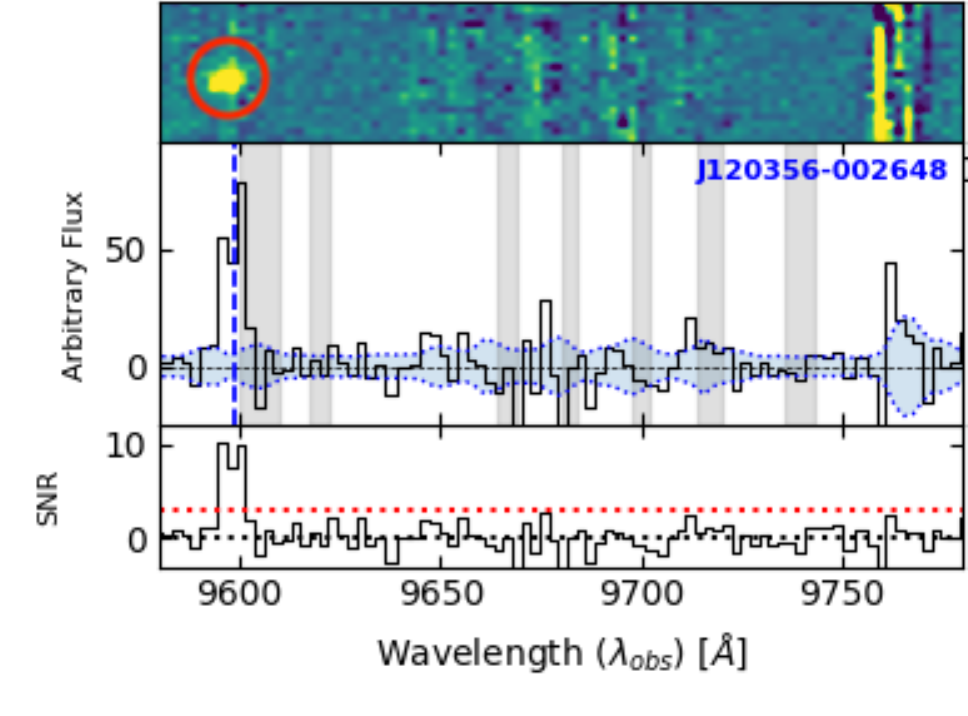}\\ \vspace{-0.5em}
		\hspace{-3.75em}\includegraphics[width=2.4in]{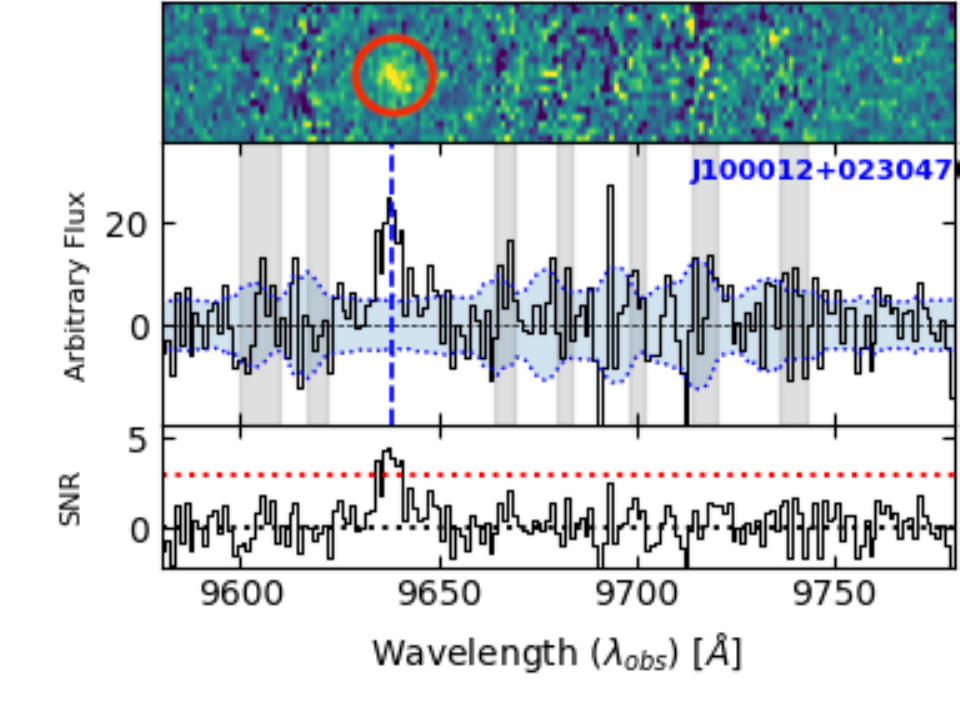} &  \hspace{-1.em}\includegraphics[width=2.4in]{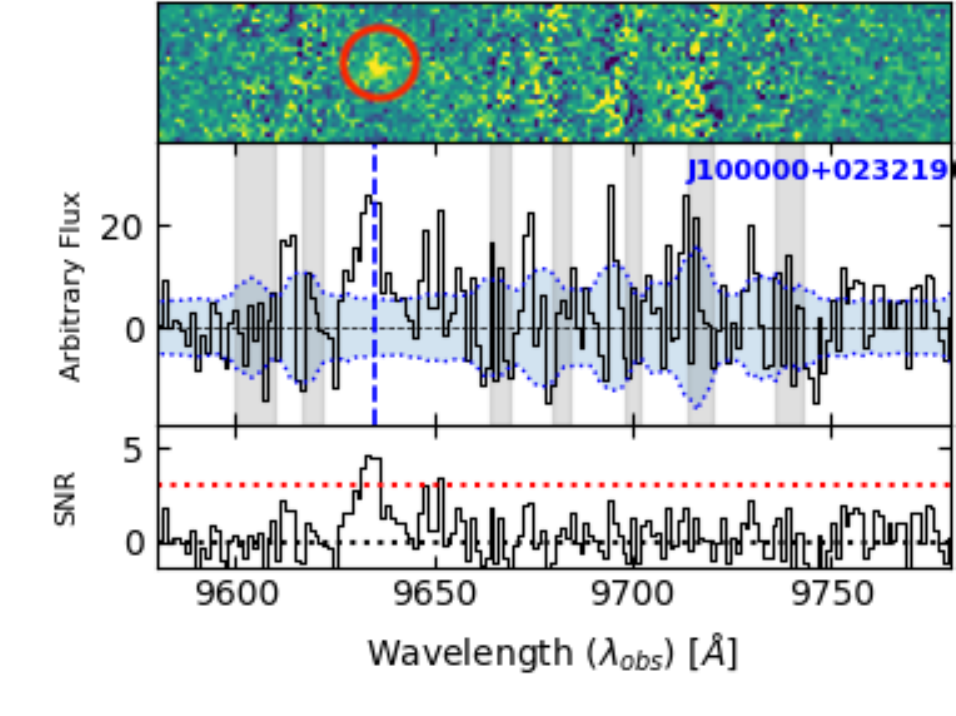} & \hspace{-1.em}\includegraphics[width=2.4in]{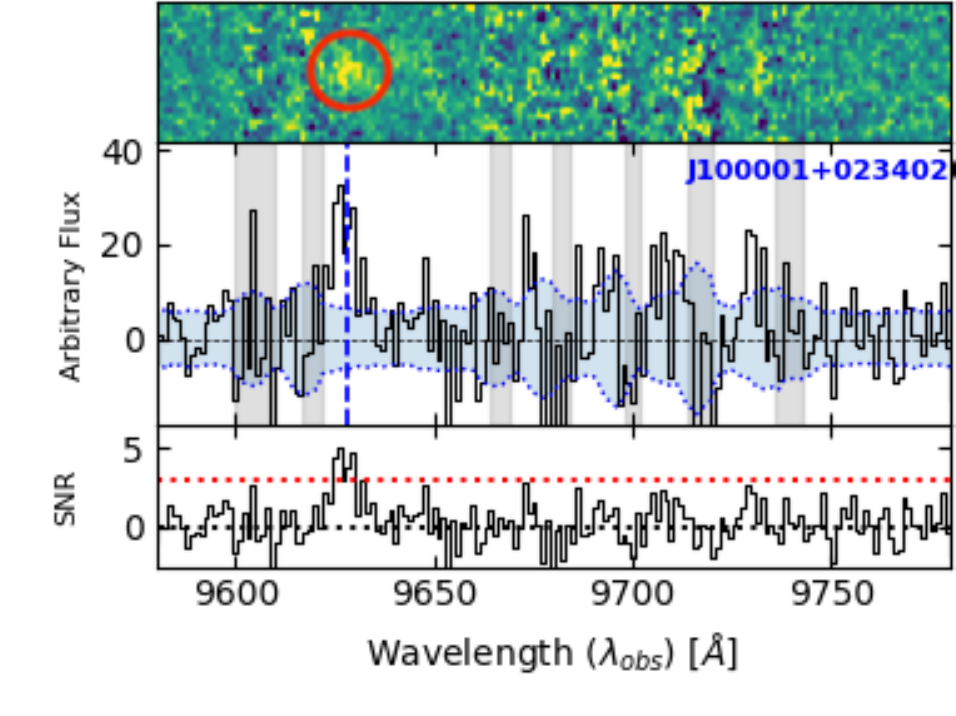}\\ \vspace{-0.5em}
		\hspace{-3.75em}\includegraphics[width=2.4in]{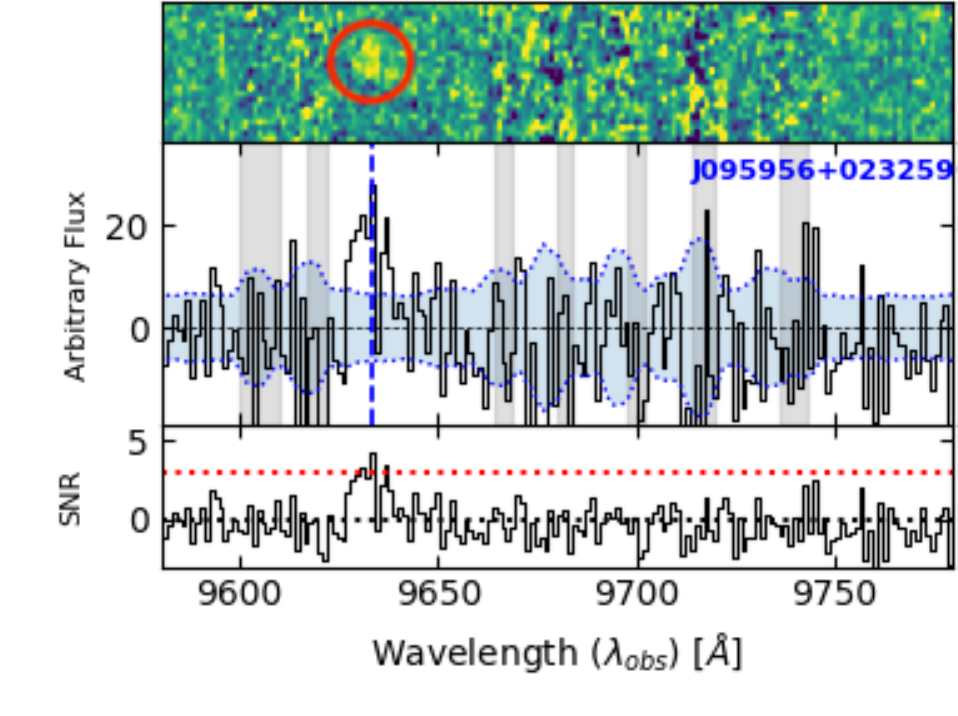} &  \hspace{-1.em}\includegraphics[width=2.4in]{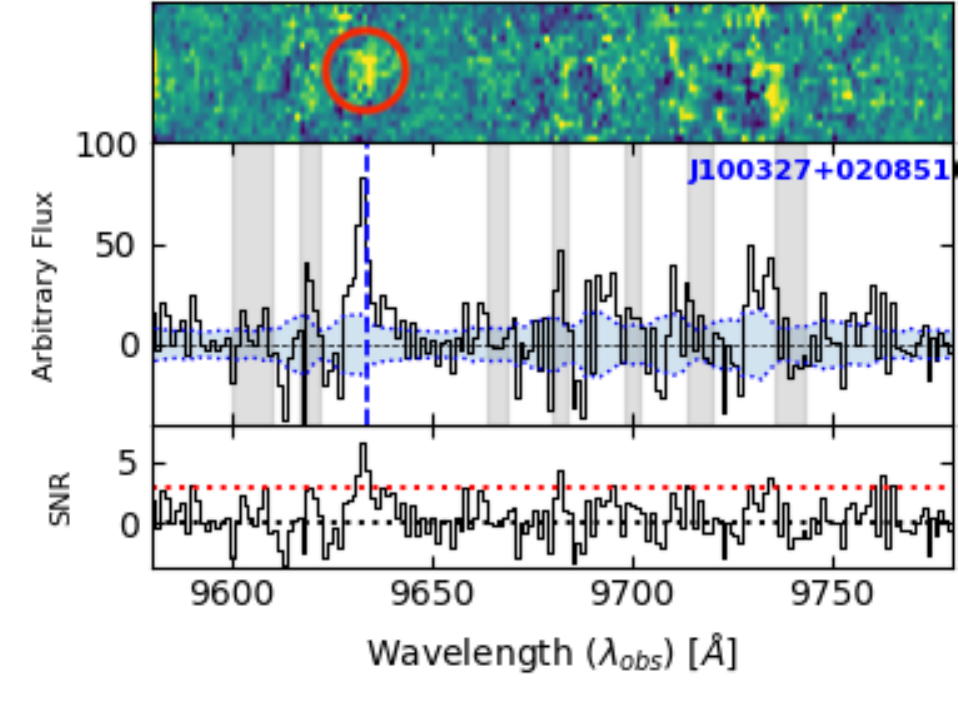} & \hspace{-1.em}\includegraphics[width=2.4in]{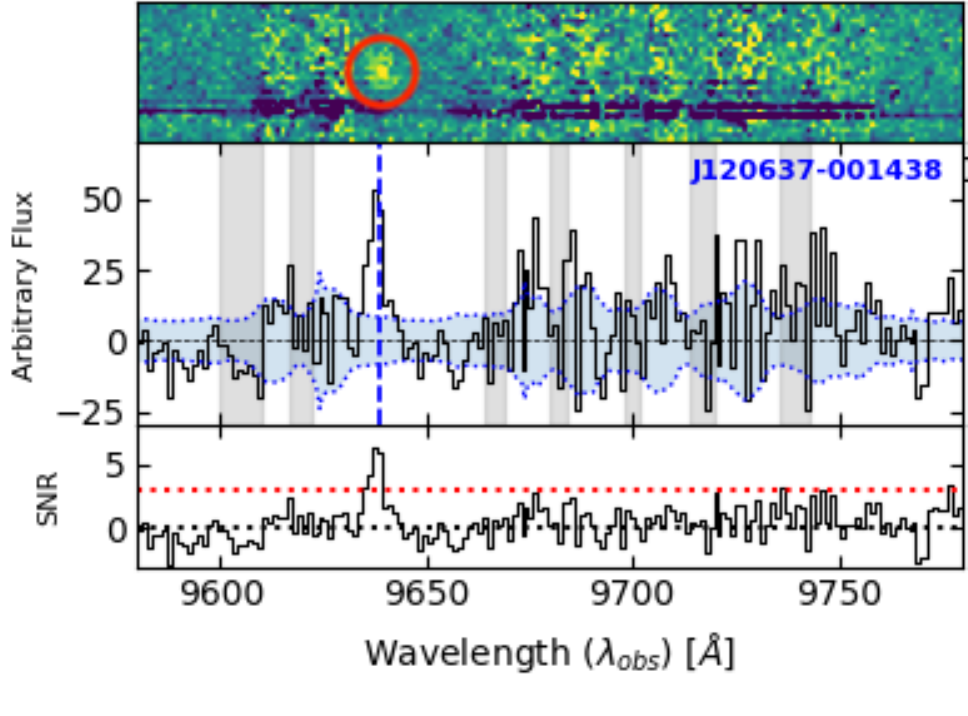}\\ \vspace{-0.5em}
		\hspace{-3.75em}\includegraphics[width=2.4in]{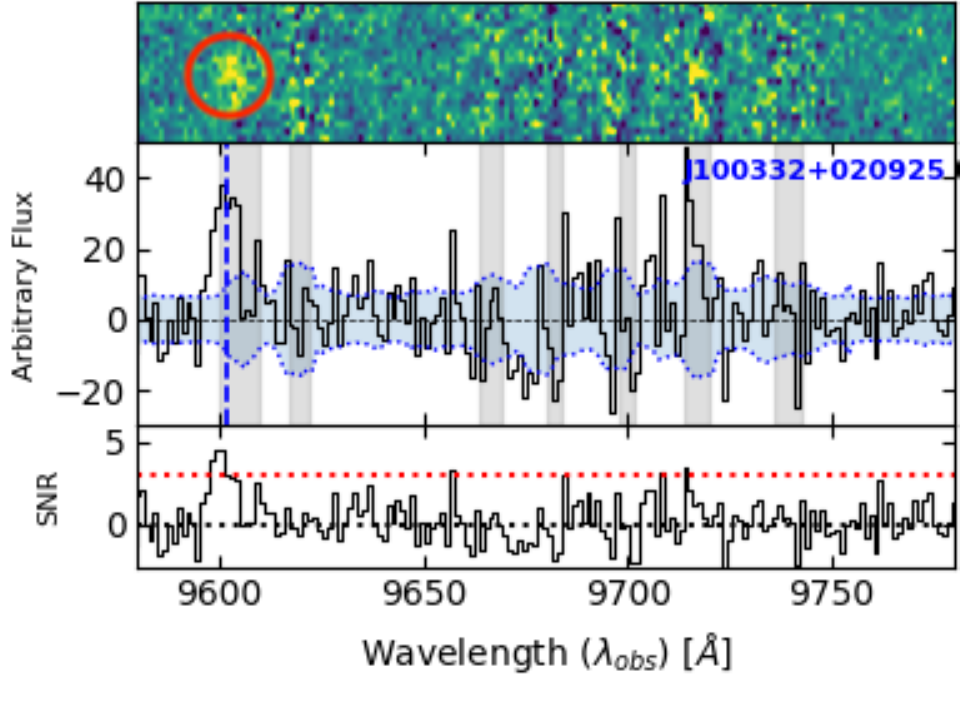} &  \hspace{-1.em}\includegraphics[width=2.4in]{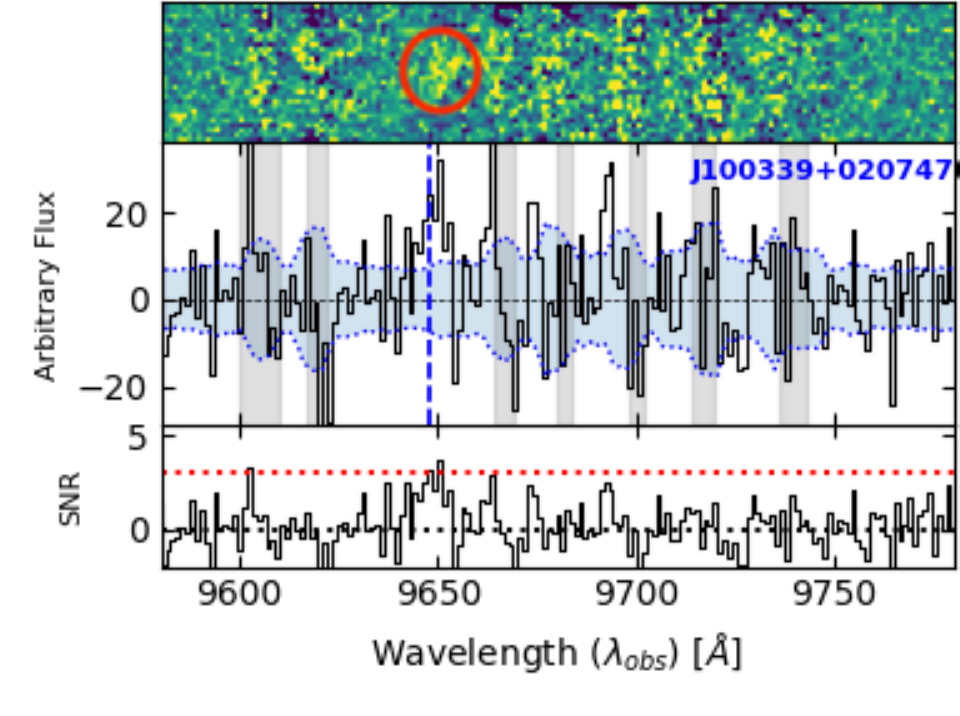} & \hspace{-1.em}\includegraphics[width=2.4in]{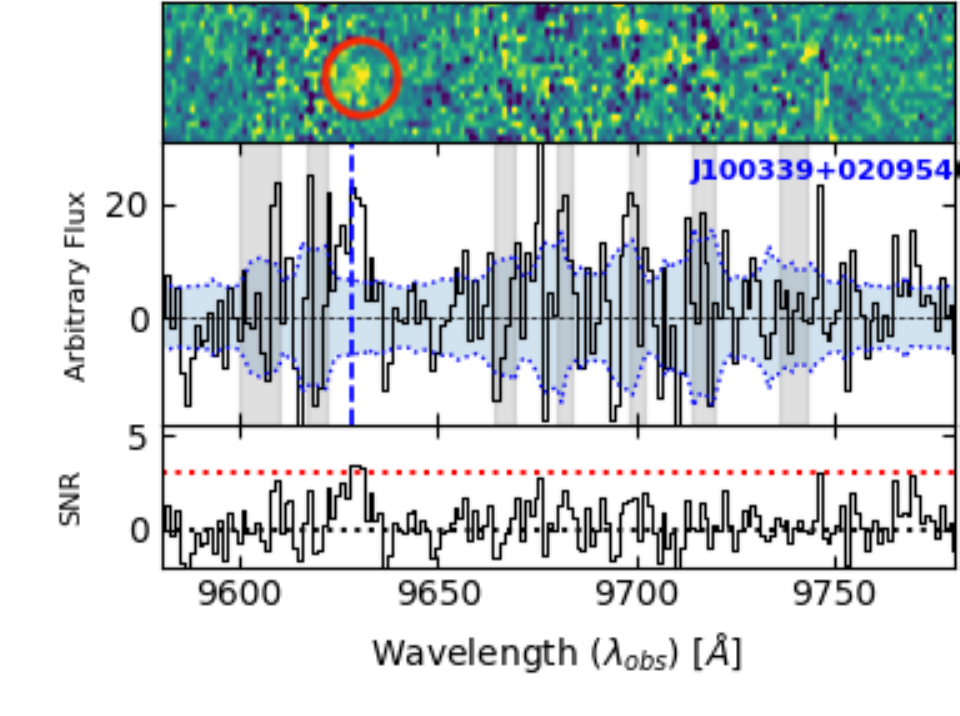}
	\end{tabular}
	\caption{LRIS spectra of confirmed LAGER LAE candidates. The top panel shows the coadded 2D spectrum with detected Ly$\alpha$ emission (\emph{red circle}). In the middle panel, the coadded 1D spectrum containing the Ly$\alpha$ line (\emph{blue line}) along with noise spectrum (\emph{blue shade}) and prominent sky emission regions (\emph{gray shade}) over-plotted are shown, and the bottom panel is the S/N spectrum with a 3$\sigma$ limit (\emph{red}) shown for reference.}
	\label{fig:lae_spec}
\end{figure*}

\begin{figure*}[h]
\begin{tabular}{ll}
	\hspace{-3.5em}\includegraphics[width=3.5in]{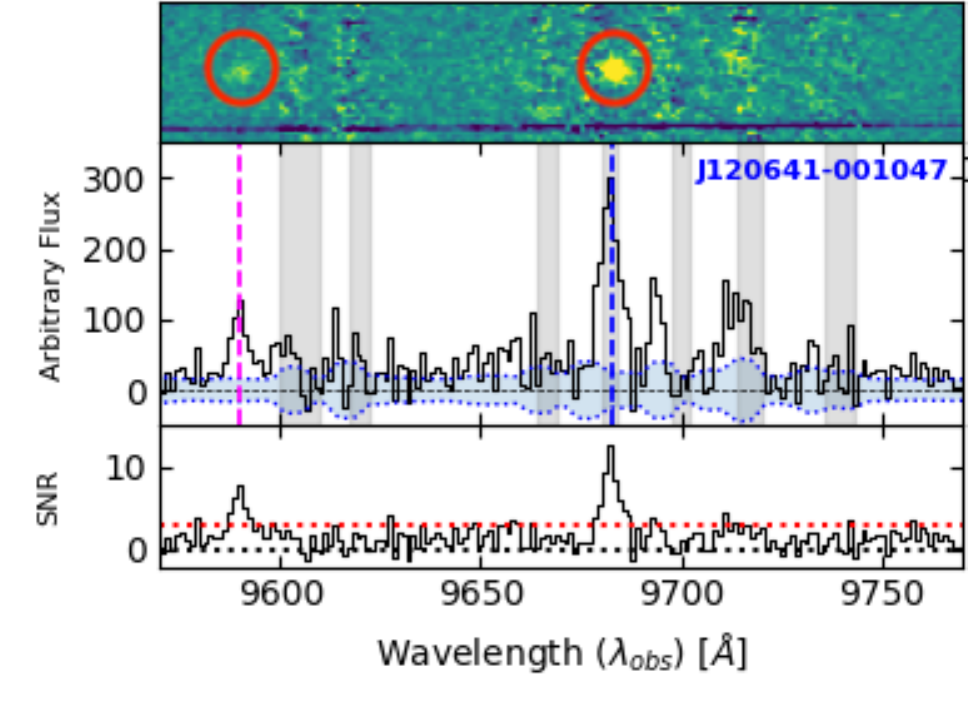} & \hspace{-1em}\includegraphics[width=3.5in]{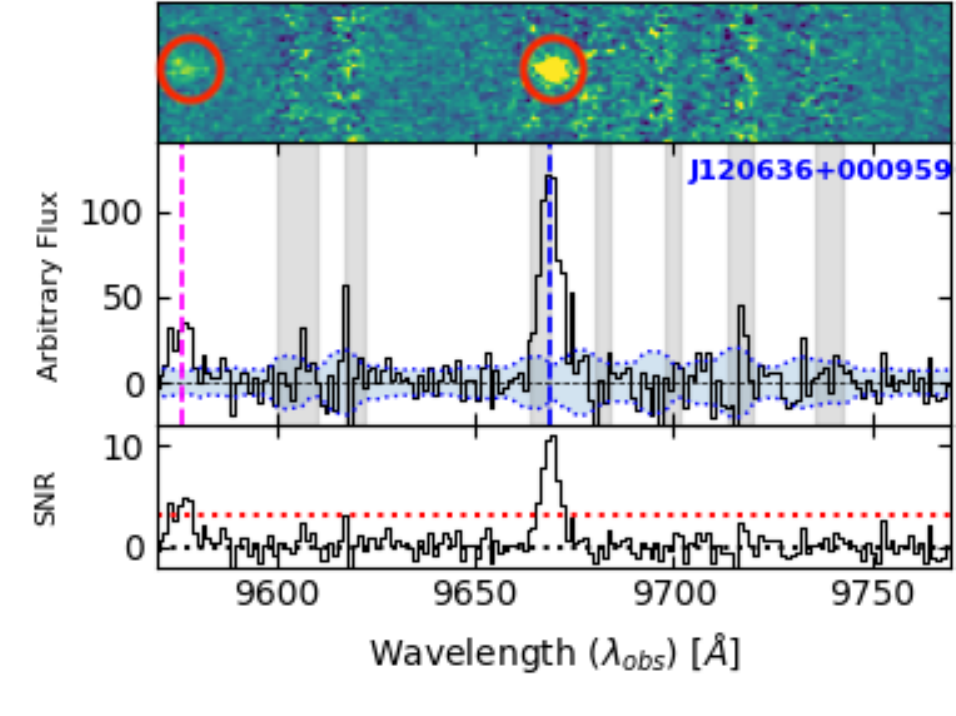} \\
\end{tabular}
\caption{LRIS spectra of two low-redshift interlopers ([O\textsc{iii}] at $z\sim0.92$). Similar to Figure \ref{fig:lae_spec}, the coadded 2D (\emph{top}), 1D (\emph{middle}) and S/N (\emph{bottom}) spectra are shown for each source. In the 1D spectrum, the detected [O\textsc{iii}]$_{\lambda4959}$ (\emph{pink line}) and [O\textsc{iii}]$_{\lambda5007}$ (\emph{blue line}) emission are also shown.}
\label{fig:oiii_spec}
\end{figure*}

Out of 21 LAE candidates across the two fields, emission lines were detected within the expected wavelength range, based on NB964 filter coverage, in 17 candidates (11 in COSMOS and 6 in WIDE-12). Only one emission line was detected in 15 of these, without any trace of continuum in any of them, whereas each of the remaining two candidates contained two emission lines within the expected wavelength range. The 2D and 1D spectra for these sources are presented in Figures \ref{fig:lae_spec} and \ref{fig:oiii_spec}.

We identified the single line in these 15 candidates as Ly$\alpha$ emission at $z \sim 6.9$ and rule out the possibility of being low-$z$ foreground emission for the following reasons: (1) None of these sources are detected in the deep \emph{gri} broad-band images reaching a 5$\sigma$ depth of $\gtrsim 27.5$ mag in COSMOS and $\gtrsim 26.5$ in WIDE-12. (2) Other prominent rest-frame optical lines can be ruled out as follows: if it were to be [O\textsc{iii}]$_{\lambda5007}$ (or [O\textsc{iii}]$_{\lambda4959}$), then we should have also detected its complementary line, [O\textsc{iii}]$_{\lambda4959}$ (or [O\textsc{iii}]$_{\lambda5007}$). If it were to be H$\alpha$, given LRIS' spectral coverage, then H$\beta$ and/or [O\textsc{iii}] should have been detected. Since the detected emission lines are narrow with no obvious double-peak profile, we rule out the possibility of any of them being the [O\textsc{ii}]$_{\lambda\lambda3727,3729}$ line. (3) Typically, Ly$\alpha$ emission from star-forming galaxies can be differentiated from other lines based on the asymmetric nature of Ly$\alpha$ line \cite[e.g.][]{Rhoads2003}. Inspection of 1D spectra in Figure \ref{fig:lae_spec} shows a hint of asymmetry in some sources (J120412-003157, J120406-002921, J095934+014406) but owing to the low spectral resolution of LRIS, most other candidates seem to show no significant asymmetry. Considering all the aforementioned evidence, we infer that the detected emission line in these 15 sources is indeed the Ly$\alpha$ line. Table \ref{tab:lager_spec} summarizes the basic properties of all confirmed LAE candidates including their spectroscopic redshifts. Three LAEs (J100332+020925, J100327+020851, J100339+020747) are part of the overdense region in COSMOS, LAGER-$z7$OD1 \citep{Hu2019} and have matching confirmations with spectroscopic follow-ups reported by \citealt{Hu2021}.

\begin{figure*}
	\includegraphics[width=\linewidth]{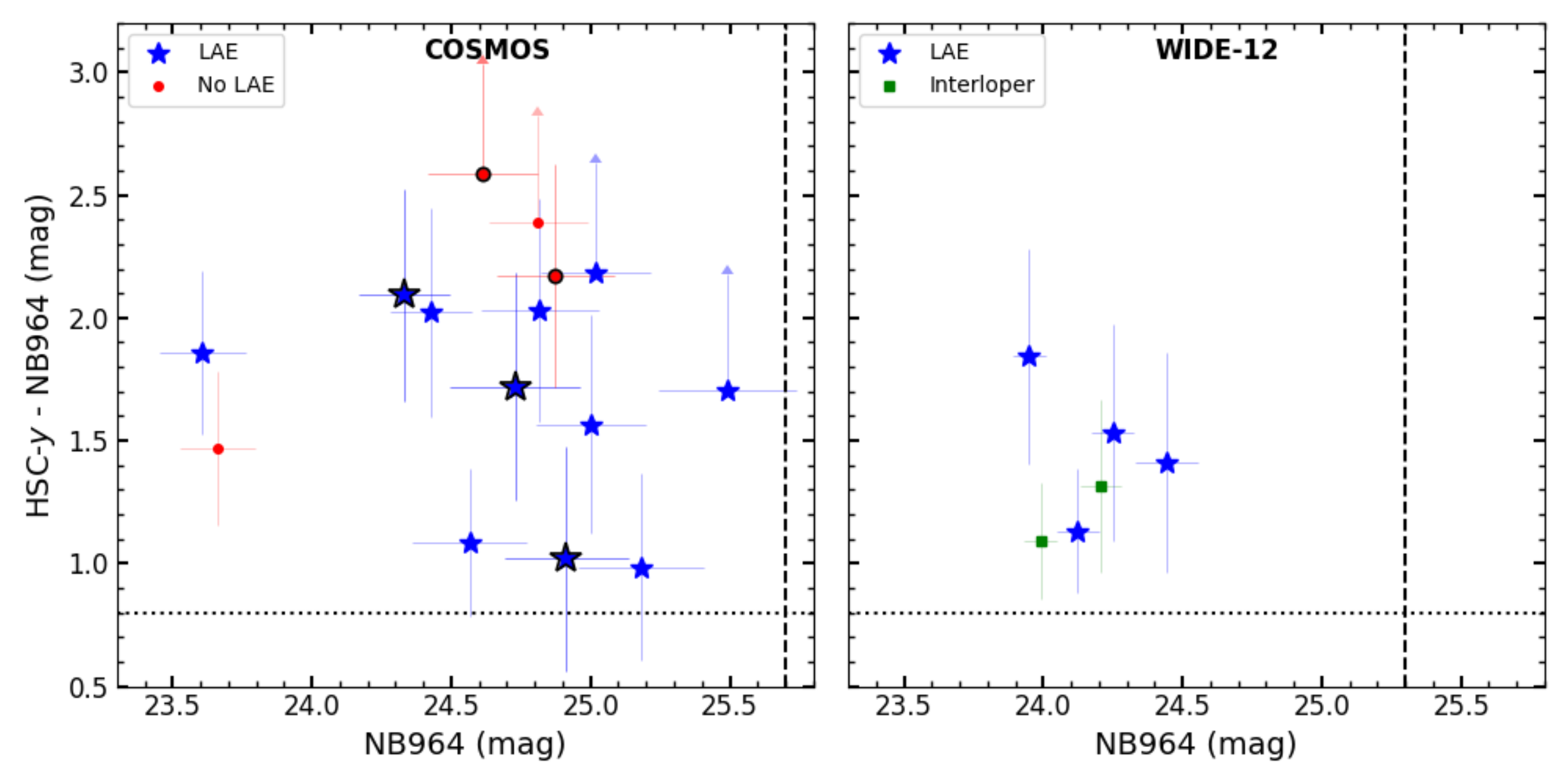}
	\caption{($y$ - NB964) color as a function of NB964 magnitude for COSMOS (\emph{left}) and WIDE-12 (\emph{right}) regions. In both panels, the color criteria used to select our LAEs \citep{Hu2019,Wold2021} are shown for reference: SNR $> 5$ (\emph{dashed line}) and $y$ -- NB964 $> 0.8$ (\emph{dotted line}). Previously confirmed LAEs by \citealt{Hu2021} are shown using \emph{black} bordered symbols.}
	\label{fig:color_mag}
\end{figure*}

Figure \ref{fig:color_mag} shows the color-magnitude diagrams of LAE candidates in COSMOS and WIDE-12 regions. Two LAEs (J120641-001047 and J120636+000959) in WIDE-12 turned out to be [O\textsc{iii}] emitters where the doublet lines, [O\textsc{iii}]$_{\lambda4959}$ and [O\textsc{iii}]$_{\lambda5007}$, were detected in both these candidates (Fig.\ \ref{fig:oiii_spec}). These two sources are not too bright in the NB color-excess ($y$ - NB964 $<$ 1.35).

No emission lines were detected in four candidates (all in COSMOS). However, two of them, J100333+020719 and J100337+020736, were spectroscopically confirmed using Magellan observations in \cite{Hu2021}. Based on their derived redshifts, the Ly$\alpha$ emission is expected at wavelengths $\sim 9624$ \AA and $\sim 9656$ \AA, respectively, which coincide with some skylines at these wavelengths. Given that the mask containing these sources had the lowest total integration times compared to other masks, we believe that the non-detections in this mask was due to these two factors combined.

\subsection{Non-detection of NV emission}
\begin{figure*}
	\includegraphics[scale=1]{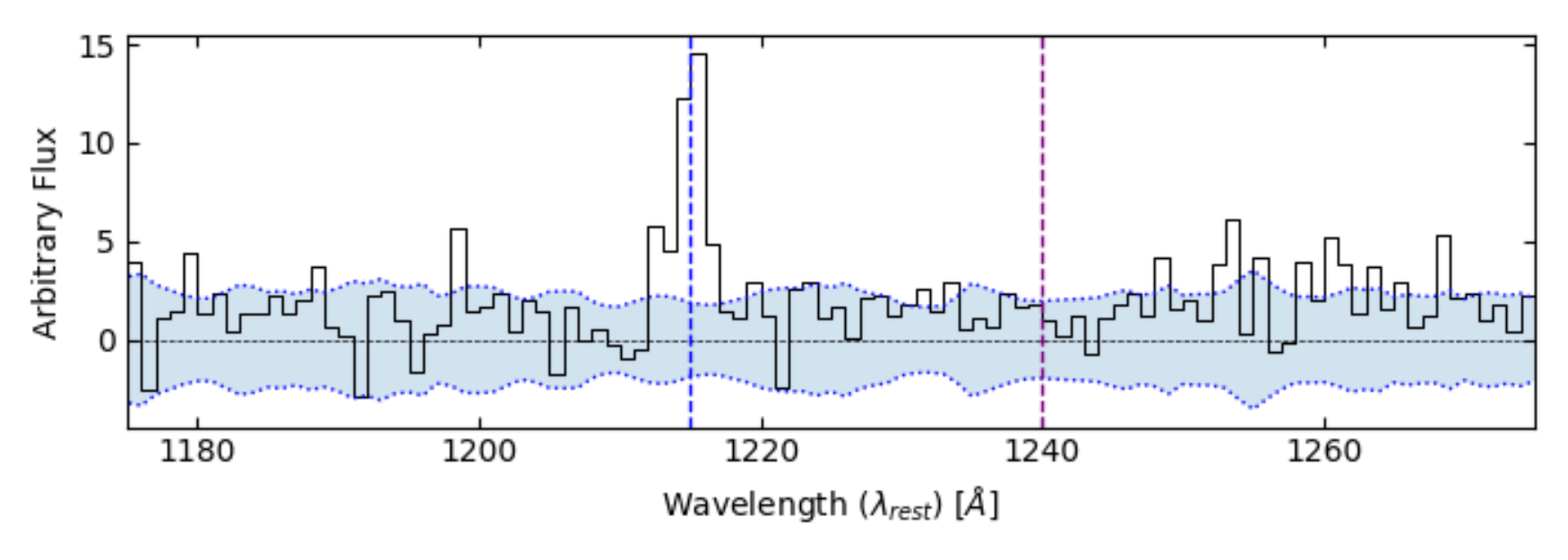}
	\caption{Inverse-variance weighted 1D stack of LAE spectra with the corresponding 1$\sigma$ error (\emph{blue shade}). No evidence is seen for N\textsc{v} emission that would be expected at a ratio of $f_{NV}/f_{Ly\alpha} > 0.3$, if the Ly$\alpha$ lines were due to AGN. The Ly$\alpha$ (\emph{blue line}) and N\textsc{v} (\emph{purple line}) wavelengths are marked in the spectrum for reference.}
	\label{fig:lae_stack}
\end{figure*}

Apart from the Ly$\alpha$ line, only other UV nebular emission line that we could detect, given the spectral coverage of LRIS, was the N\textsc{v}$_{\lambda1240}$. Recent works involving $z \gtrsim 7$ LAE search have found (tentative) N\textsc{v} detections \cite[e.g.][]{Tilvi2016,Hu2017} indicating the presence of a (weak) AGN and possibly metal-rich gas. However, as mentioned earlier, none of our confirmed LAE sources contain any other emission line except Ly$\alpha$.  Therefore, we performed an inverse-variance weighted stack of 1D spectra of all our LAEs to look for N\textsc{v} emission but we found no statistically-significant detection at the expected wavelength (Figure \ref{fig:lae_stack}). The 2$\sigma$ upper-limit on the flux ratio of N\textsc{v}/Ly$\alpha$ is $f_{NV}/f_{Ly\alpha} \lesssim 0.27$, which is consistent with recent findings from studies targeting bright $z\gtrsim6$ LAEs ($f_{NV}/f_{Ly\alpha} < 0.3$; \citealt{Mainali2018,Shibuya2018,Yang2019}).

\section{Summary} \label{sec:summary}
We report spectroscopic confirmation of 15 LAE candidates from the LAGER survey based on observations using the Keck/LRIS spectrograph. Four of these are the first spectroscopic confirmations of LAE candidates from the WIDE-12 sample. Of the 11 confirmed LAEs in COSMOS, 8 are new confirmations and 3 have matching confirmations with \cite{Hu2021} which is part of the overdense region, LAGER-$z7$OD1. Overall, including two additional confirmations from previous spectroscopic follow-up using Magellan \citep{Hu2021}, the success rate of LAE confirmations is $\sim81$\%. We found no significant trace of N\textsc{v} emission, an indicator of AGN activity, in any of these sources. Considering previous LAE confirmations \citep{Hu2017,Yang2019,Hu2021} and those presented in this work, 33 unique LAE sources from LAGER are now spectroscopically confirmed which has more than doubled the sample of spectroscopically confirmed LAE sources at $z\sim7$.

\newpage
\begin{acknowledgments}
We thank NASA for its support to ASU via contract NNG16PJ33C, "Studying Cosmic Dawn with WFIRST". IGBW is supported by an appointment to the NASA Postdoctoral Program at the Goddard Space Flight Center, administered by the Universities Space Research Association through a contract with NASA. The material is based upon work supported by NASA under award number 80GSFC21M0002. JW thanks support from National Natural Science Foundation of China (grant Nos. 11890693 \& 12033006) and the CAS Frontier Science Key Research Program (QYZDJ-SSW-SLH006). ZYZ acknowledges support by the National Science Foundation of China (11773051, 12022303), the China-Chile Joint Research Fund and the CAS Pioneer Hundred Talents Program. 

The data presented herein were obtained at the W. M. Keck Observatory, which is operated as a scientific partnership among the California Institute of Technology, the University of California and the National Aeronautics and Space Administration. The Observatory was made possible by the generous financial support of the W. M. Keck Foundation. The authors wish to recognize and acknowledge the very significant cultural role and reverence that the summit of Maunakea has always had within the indigenous Hawaiian community.  We are most fortunate to have the opportunity to conduct observations from this mountain.
\end{acknowledgments}

\bibliography{refs}{}
\bibliographystyle{aasjournal}



\end{document}